\documentstyle[aps,preprint,epsf,floats]{revtex}
\tighten

\def\erf{{\rm erf}}
\def\eq{{\rm eq}}
\def\for{{\rm for}}
\def\i{{\rm i}}
\def\min{{\rm min}}
\def\m{{\rm m}}
\def\max{{\rm max}}
\def\P{{\rm P}}
\def\r{{\rm r}}
\def\reh{{\rm reh}}
\def\rmd{{\rm d}}
\def\rme{{\rm e}}
\def\rmint{{\rm int}}
\def\sgn{{\rm sgn}}
\def\therm{{\rm therm}}
\def\w{{\rm w}}
\def\where{{\rm where}}
           
\begin{document}
\draft
\preprint{\small OUTP-01-28P}
\title{No cosmological domain wall problem for weakly coupled fields}
\author{Horacio Casini and Subir Sarkar}
\address{Theoretical Physics, University of Oxford, 1 Keble Road, 
         Oxford OX1 3NP, UK}
\bigskip
\maketitle
\begin{abstract}
After inflation occurs, a weakly coupled scalar field will in general
not be in thermal equilibrium but have a distribution of values
determined by the inflationary Hubble parameter. If such a field
subsequently undergoes discrete symmetry breaking, then the different
degenerate vacua may not be equally populated so the domain walls
which form will be `biased' and the wall network will subsequently
collapse. Thus the cosmological domain wall problem may be solved for
sufficiently weakly coupled fields in a post-inflationary universe. We
quantify the criteria for determining whether this does happen, using
a Higgs-like potential with a spontaneously broken $Z_2$ symmetry.
\end{abstract}
\bigskip
\widetext
\small

\section{Introduction}
\label{intro}

It is generally believed that if {\em discrete} symmetries of scalar
fields are spontaneously broken as the universe cools down, then there
would be severe difficulties for its subsequent evolution
\cite{zko75}. This is because topological defects --- domain walls ---
would form at the boundaries of the different degenerate vacua chosen
in causally disconnected regions following the symmetry breaking phase
transition \cite{kib76} and would eventually come to dominate the
total energy density, in conflict with observations \cite{vs94}. To
avoid this requires the energy scale of the symmetry breaking phase
transition to be lower than $\sim100$~MeV; in fact it must be less
than $\sim1$~MeV if the anisotropy induced by the walls in the cosmic
microwave background radiation is to be below experimental limits
\cite{zko75,vs94}. This is a severe constraint on attempts to extend
physics beyond the Standard Model, which often involve introducing
such discrete symmetries \cite{discrete}.

In some special circumstances, the broken discrete symmetry may not be
restored at high temperature so domain walls would never form
\cite{symnonrestor}. There would also appear to be no problem if the
symmetry breaking occurs prior to inflation since one would then
expect the density of any resulting topological defects to be
exponentially diluted away. However defects can still form in this
case through quantum fluctuations of the scalar field {\em during}
inflation \cite{ll90,hp91,ny92} if the mass of the field is less than
the inflationary Hubble parameter. When inflation is driven by a
F-term supergravity potential, the breaking of supersymmetry by the
large vacuum energy gives all scalar fields, including the inflaton
itself, a mass-squared of ${\cal O}(H^2)$ \cite{sugra1,sugra2}; in
this case fluctuations are negligible and walls will {\em not}
form. However if we consider instead e.g. a D-term or no-scale
inflationary potential \cite{lr99}, the scalar field may remain light
relative to the Hubble parameter so the above mechanism will be
operative and domain walls will form.

Although the first such paper \cite{ll90} considered an axion field,
subsequent work \cite{hp91,ny92} has been mainly concerned with scalar
fields which have sufficiently strong couplings that the vacuum
expectation value (vev) during inflation does not increase much above
the Hubble parameter. The field then remains uncorrelated on spatial
scales larger than the Hubble radius and defects form during or at the
end of inflation. However very weakly coupled scalar fields are
arguably of more interest in cosmology. For example the field
responsible for driving inflation should have very small couplings in
order that its quantum fluctuations not contribute excessively to the
anisotropy of the cosmic microwave background \cite{lr99}. There has
been much interest in `quintessence' \cite{bin00} --- a very weakly
coupled evolving field that may account for the tiny vacuum energy
that is suggested by some astronomical observations. Weakly coupled
fields can also be a source of dark matter through their coherent
oscillations \cite{peb00}. In a recent paper \cite{dgs01} an extremely
weakly coupled dilaton field that forms domain walls is proposed as a
way of binding the matter in spiral galaxies and producing their
characteristic flat rotation curves (as an alternative to cold dark
matter). Particularly in the context of this model, it is interesting
to ask whether the above mechanism would indeed create stable domain
walls.

The point is that such a very weakly coupled field will be {\em
correlated} on super-horizon scales at the end of inflation and not be
brought back into thermal equilibrium during the reheating process
since it has no couplings to the thermal plasma or to the
inflaton. The field will oscillate coherently during the
post-inflationary Friedman-Lemaitre-Robertson-Walker (FLRW) expansion
era and when the expansion redshift reduces the energy in its coherent
oscillations it will settle into different symmetry-breaking vacua on
spatial scales larger than the Hubble radius, thus forming defects. It
would be likely for the same vacuum to be chosen in different
(apparently causally disconnected) regions. A `bias' could thus be
generated in the probabilities for populating the distinct vacuua even
if they are energetically degenerate \cite{lto93}. After the walls
form, such a bias, even if very small, will result in exponential
decay of the wall network, as has been demonstrated both analytically
and numerically \cite{clo96,hin96,lsw97}. Thus there may be no domain
wall problem for weakly coupled fields in a post-inflationary
universe.

Our aim is to quantify the bias that would be created for such a
hypothetical field with specified properties in order to determine the
fate of the domain walls formed. We consider the problem in its
simplest form and focus on domain wall formation through inflationary
fluctuations in a spontaneously broken $Z_2$ theory of a real scalar
field with the Higgs-like potential
\begin{equation}
 V (\phi) = -\frac{1}{2}m^{2}\phi^2 + \frac{1}{4}\lambda\phi^4 
          = \frac{\lambda}{4}\left(\phi^2 - v^2\right)^2\,,
\label{poti}
\end{equation}
where $v=m/\sqrt{\lambda}$. First we study (Section~\ref{stoc}) the
stochastic evolution of the field perturbations during the
inflationary epoch which set the relevant initial conditions. In
Section~\ref{evol} we follow the evolution of the fluctuations during
the FLRW expansion until the field drops into its potential minima and
the domain walls are formed. The bias between the degenerate vacua is
calculated in Section~{\ref{bia}. We review the history of the field
evolution in Section~\ref{hist} and identify the regions in the
parameter space of the above model where domain walls do not
survive. Finally we present our conclusions and comment on specific
models concerning domain walls such as Ref.\cite{dgs01}.

\section{Stochastic approach for the initial conditions}
\label{stoc}

During inflation the smooth component of a slowly evolving scalar
field can be considered (on scales larger than the horizon) to be a
classical variable subject to stochastic noise (contributed by the
field modes whose exponentially increasing wavelength causes them to
`exit the horizon', becoming part of the coarse-grained field)
\cite{sta82,gl87,rey87,more,lin90}. The Langevin equation governing
the coarse-grained field $\phi$ is \cite{sta82}
\begin{equation}
 \dot{\phi} = - \frac{V^\prime(\phi)}{3H_\i} 
              + \frac{H_\i^{3/2}}{2\pi} \eta (t)\,,  
\label{lang}
\end{equation}
where the white noise $\eta$ satisfies
\begin{equation}
 \left\langle\eta (t) \eta(t^\prime)\right\rangle = \delta (t-t^\prime)\,.
\label{noise}
\end{equation}
This equation can be restated as a Fokker-Plank equation for the
probability distribution $P(\phi,t)$ of the field values in a given
coarse-grained domain \cite{sta82}:
\begin{equation}
 \frac{\partial P (\phi, t)}{\partial t} = 
 \frac{\partial}{\partial\phi} 
 \left(\frac{1}{3H_\i}\frac{\partial V}{\partial \phi} P(\phi, t)\right) 
 + \frac{H_\i^3}{8\pi^2}\frac{\partial^2 P(\phi,t)}{\partial\phi^2}\,.
\label{fp}
\end{equation}
Here $H_\i$, the Hubble parameter during inflation, is taken to be
independent of $\phi$, i.e this field is assumed {\em not} to
contribute significantly to the vacuum energy during inflation. (In
the analogous equation for the inflaton field, $H_\i$ is itself a
function of $\phi$ giving rise to ordering ambiguities in the
corresponding Fokker-Planck equation.)

If the field evolution is `slow-roll', then the force term (involving
the potential) can be neglected compared with the noise term in
Eq.(\ref{fp}). Then starting from a given value of the field
$\phi=\bar{\phi}$ averaged over a patch of size $H_\i^{-1}$ when
cosmologically relevant scales `exit the horizon', the solution to
this equation is the Gaussian distribution:
\begin{equation}
 P_\phi \equiv P (\phi, \bar{\phi}, t) = \sqrt{\frac{8\pi^3}{H_\i^3 t}}\,  
 \exp{\left[-\frac{2\pi^2\left(\phi-\bar{\phi}\right)^2}{H_\i^3 t}\right]},
\end{equation}
where $H_\i$ is taken to be approximately constant as is required for
successful inflation \cite{lr99}.

The mean value $\left\langle\phi^2\right\rangle=H_\i^{3}t/4\pi^2$
grows linearly with time as in Brownian motion \cite{fluc}. Relating
the time during inflation to the cosmological scale through
$l^{-1}\sim\,H_\i\rme^{-H_{\i}t}$, we can write the probability
distribution as \cite{ll90,lto93}
\begin{equation}
 P_\phi (\phi, \bar\phi) = \frac{1}{\sqrt{2\pi}\sigma}
  \exp\left[-\frac{1}{2\sigma^2}\left(\phi - \bar{\phi}\right)^2\right]\,,
\label{prob}
\end{equation}
where $\sigma^2$ is the quadratic dispersion of the field. This can be
obtained, with reference to the noise term in Eq.(\ref{lang}), as the
sum of independent Gaussian distributions with dispersion-squared
$(H_\i/2\pi)^2$, one for each e-fold of inflation. The sum is to be
taken over the period when scales between $l_\min$ and $l_\max$ leave
the horizon, where $l_\min$ corresponds to the Hubble radius at some
moment during the FLRW era (representing an ultraviolet cutoff, given
that we are interested in the super-horizon behavior) and $l_\max$ is
the biggest spatial scale of interest, i.e. of order the present
Hubble radius $H_0^{-1}$. The total dispersion-squared is then just
the sum of the dispersion-squared for the independent probabilities:
\begin{equation}
 \sigma^2 = \frac{H_\i^2}{4\pi^2} \int_{l_\min}^{l_\max} \rmd \log l
          = \frac{H_\i^2}{4\pi^2} \log\left(\frac{H_0^{-1}}{l_\min}\right)\,,
\label{ttt}
\end{equation}
i.e. of ${\cal O}(H_\i^2)$ in the cases of interest.

The formula (\ref{prob}) assumes that the value of the force term in
Eq.(\ref{lang}) is negligible in comparison with the noise term so the
value of $\bar{\phi}$ is not determined. However, if inflation
continues for a large number of e-folds the force term will impede the
tendency of the distribution to widen indefinitely. In this case
stochastic equilibrium is achieved and it is possible to give a
probabilistic prediction for the initial $\bar{\phi}$ using the {\em
stationary} solution for the Fokker-Planck equation (while
Eq.(\ref{prob}) still gives the distribution for the field at the end
of inflation on cosmologically interesting scales).\footnote{This has
been shown in studies of `eternal inflation' \cite{eternal}.}

The stationary case ${\partial P}/{\partial t}=0$ can be solved to
obtain the probability distribution of the averaged field $\bar\phi$:
\begin{equation}
 P_{\bar\phi} = 
   C_1 \exp{\left(-\frac{8\pi^2}{3}\frac{V(\bar\phi)}{H_\i^4}\right)} 
   + C_{2} \exp{\left(-\frac{8\pi^2}{3}\frac{V(\bar\phi)}{H_\i^4}\right)}
     \int_{0}^{\bar\phi}\exp{\left(\frac{8\pi^2}{3}
     \frac{V(\phi^\prime)}{H_\i^4}\right)} \rmd\phi^{\prime}\,.
\end{equation}
If the potential is an even function the first term is also even while
the second term is odd. Therefore, if the potential remains positive
for large $\phi$, the second term will have greater absolute values
than the first at some point, and as it is an odd function the
probability would have negative values. This shows that the second
term is unphysical. Thus, in the stationary case the normalized
probability distribution is just:
\begin{equation}
 P_{\bar\phi} = 
  \exp{\left(-\frac{8\pi^2}{3}\frac{V(\bar\phi)}{H_\i^4}\right)}
  \left/ \int_{-\infty}^{+\infty} \exp{\left(-\frac{8\pi^2}{3}
  \frac{V (\phi^\prime)}{H_\i^4}\right)}\rmd\phi^\prime\right.\,.
\end{equation}
Defining the dimensionless normalized fields $\chi\equiv\phi/v$, and
$\bar\chi\equiv\bar\phi/v$, this is a one-parameter function for our chosen
potential (\ref{poti}):
\begin{equation}
 P_{\bar\chi} = \frac{2}{\pi\left[I(\frac{1}{4},
  \frac{\pi^2}{3\beta^4}) +
  I(-\frac{1}{4},\frac{\pi^2}{3\beta^4})\right]}
  \exp{\left[-\frac{2\pi^2}{3\beta^4}\left(\bar{\chi}^4 -
  2\bar{\chi}^2 + \frac{1}{2}\right)\right]}, \quad 
 \beta \equiv \lambda^{-1/4}\frac{H_\i}{v}\,,
\label{dis}
\end{equation}
where $I(x,y)$ is the modified Bessel function of first kind. This is
plotted in Fig.\ref{ppro} for various values of $\beta$.

\begin{figure}[htb]
\hspace{3cm}\epsfxsize10cm\epsffile{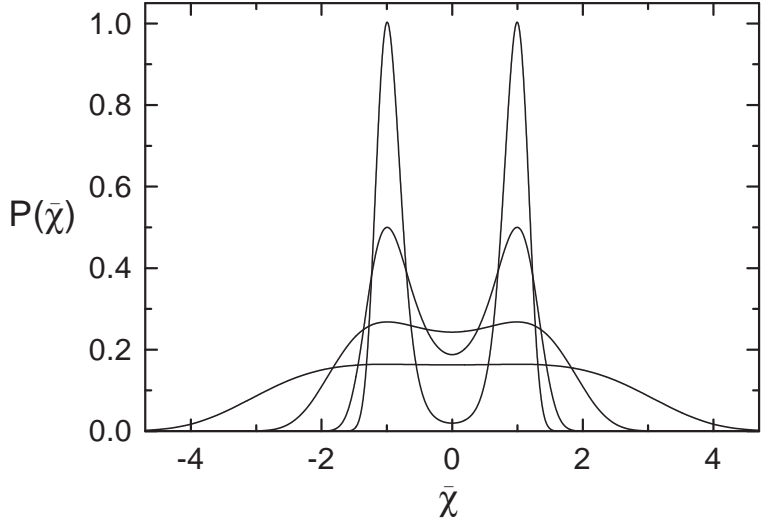}
\bigskip
\caption{Stochastic equilibrium probability distribution for the
(normalized) field vev during inflation. The curves shown are for
$\beta=1.2, 1.6, 2.8, 5$ in order of decreasing height.}
\label{ppro}
\end{figure}

Since the height of the potential barrier at the origin separating the
two vacua is $h^4=\case{1}{4}\lambda\,v^4$ from Eq.(\ref{poti}), the
parameter $\beta$ in Eq.(\ref{dis}) is just
$\beta=H_\i/\sqrt{2}h$. Thus for small $\beta$ the probability of
transitions between the potential wells is exponentially suppressed;
the discrete symmetry is broken during inflation and the field is
localized in one minimum or the other so domain walls will not form
(assuming that the symmetry is not restored during reheating after
inflation). However when $H_\i\gg\,h$ the fluctuations are large
enough to make the probability distribution flat as seen in
Fig.\ref{ppro}.

In general the average value of the potential energy in the field
$\phi$ is
\begin{equation}
 \left\langle V \right\rangle 
 = \int V (\bar{\phi}) P_{\bar\phi}\, \rmd\bar{\phi} 
 = g (\beta) H_\i^4\,,
\end{equation}
where $g(\beta)$ is a smooth positive function with maximum value
$0.02$ and limiting values $1.9\times10^{-2}$ for $\beta\rightarrow0$
and $9.4\times10^{-3}$ for $\beta\rightarrow\infty$. Thus the energy
density of the scalar field under consideration is indeed negligible
when compared with that of the inflaton $V_\i\simeq3H_\i^{2}M_\P^2$,
as was implicitly assumed in neglecting the $\phi$ dependence of
$H_\i$. (Here $M_\P\equiv1/\sqrt{8\pi\,G}\simeq2.4\times10^{18}$~GeV
is the normalized Planck scale.) Note that the COBE observations of
large-scale anisotropy in the cosmic microwave background set a strict
upper bound on the Hubble parameter during inflation \cite{lr99}
\begin{equation}
 V_\i^{1/4} \ll 0.027 M_\P \Longrightarrow 
 \frac{H_\i}{M_\P} \ll 4.2 \times 10^{-4}.
\label{scale}
\end{equation}
This still allows the energy scale of inflation to be as high as the
GUT scale ($\sim10^{16}$~GeV) but in specific models it can be much
lower than this, in particular in `new' inflationary models with a
quadratic leading term in the potential \cite{lowscale}.

The average value of $|\bar{\phi}|$ is just $v$ for small $\beta$,
while for large $\beta$ it is $\sim0.305\,H_\i/\lambda^{1/4}$. Thus if
$\lambda\lesssim8\times10^{-3}$, the field vev grows above $H_\i$. The
number of e-folds of inflation that are necessary to achieve the
stationary distribution in this case must exceed
$(\langle\bar{\phi}\rangle/\delta\phi)^2\simeq\lambda^{-1/2}$, taking
the field increment per e-fold to be
$\delta\phi\simeq\,H_\i/2\pi$. Otherwise one cannot predict an unique
probability distribution for the field.

The slow-roll condition $\ddot{\phi}\ll\,3H_\i\dot{\phi}$ implies
$|V^{\prime\prime}(\phi)|\ll9H_\i^2$. For the potential (\ref{poti})
and the distribution (\ref{dis}) this translates into
$m^2\ll\case{9}{2}H_\i^2$ if the quadratic term in the potential
dominates (small $\beta$). When the quartic term dominates (large
$\beta$) one must require $\lambda\lesssim1$. If the slow-roll conditions
are not obeyed, the classical stochastic treatment does not apply and
the quantum creation of particles is exponentially suppressed
\cite{argen}. For the case under consideration, $h \ll H_\i$, the relevant slow-roll condition is $\lambda\lesssim1$ and these two conditions together imply $m\ll\,H_\i$.

\section{The evolution during the FLRW era}
\label{evol}

In Ref.\cite{lto93} it was shown that the probability distribution for
the coarse-grained field outside the horizon during the FLRW era
following the inflationary era evolves into a non-Gaussian
distribution. We are interested here in obtaining a discrete
probability distribution for the two vacua of the field. We follow
the evolution of the field from its initial value $\phi_o$ at the end
of inflation until it settles down into one of its discrete minima. By
obtaining the distribution function $f(\phi_o)$ whose values are
$\pm1$ according to the vacuum finally chosen, we can compute the bias
between the vacua.

The equation of motion for the modes of a scalar field outside the horizon
during the FLRW era is
\begin{equation}
 \frac{\rmd^2\phi}{\rmd t^2} + 3\frac{c}{t}\frac{\rmd\phi}{\rmd t} + 
 V^\prime = 0\,, \quad \where \quad c = Ht\,.
\label{hubble}
\end{equation}
We have rewritten the Hubble parameter in terms of the variable
$c$ which equals $1/2\,(2/3)$ for a radiation- (matter-) dominated
universe. The initial conditions are $c/t_0=H_\i$, $\phi=\phi_o$, and
$\dot{\phi}_0=0$. In general there is no exact solution for this
equation, however a good approximation may easily be obtained
\cite{tur83}.

Let us first suppose that the friction term is relatively unimportant
so at first approximation the field oscillates in its potential
minimum at the end of inflation according to the Lagrangian
$L=\case{1}{2}\dot{\phi}^2-V$. Then energy conservation implies
\begin{equation}
 \dot{\phi}^2 = 2 (V_\max - V)\,,
\end{equation}
where $V_\max$ is the maximum value of the potential energy during the
oscillation. Thus the oscillation period is given (for a symmetric
potential) by
\begin{equation}
 \Delta t = 4\int_{0}^{\phi_\max} \frac{\rmd\phi}{\sqrt{2(V_\max - V)}}\,.
\label{periodo}
\end{equation}
When $H\ll\omega=2\pi/\Delta\,t$ the friction term in
Eq.(\ref{hubble}) is $\omega/H$ times smaller than the other
terms so the assumption of negligible friction is valid. Until $H$
drops below $\omega$ the field remains approximately fixed since the
friction is relatively high and the dynamical time exceeds the
expansion age $H^{-1}$. After this point the field starts oscillating,
losing energy density of order
\begin{equation}
 \Delta\rho = \oint 3H \dot{\phi}\,\rmd\phi = 
 12 H \int_{0}^{\phi_\max} \sqrt{2(V_\max - V)} \rmd\phi
\label{deltaro}
\end{equation}
in each oscillation. Here we have used the fact that in this regime $H$
is approximately constant during each oscillation period. 

For future use we also consider the case of a general power law
potential 
\begin{equation}
 V (\phi) = \frac{\lambda}{\gamma} \phi^\gamma \,. 
\label{genpot}
\end{equation}
Let $\phi_n$ be the values of $\phi$ that are turning points of the
trajectory, $t_n$ the corresponding times and $\rho_n$ the
corresponding energy densities. Evidently $\phi_n\rightarrow0$ and
$t_n$ grows, eventually going to infinity. Equations (\ref{periodo})
and (\ref{deltaro}) imply the following relations between these
quantities
\begin{eqnarray}
 \Delta\phi_n &=& \frac{\Delta\rho_n}{\lambda\phi_n^{\gamma-1}} =
 -\frac{k_2}{t_n}\phi_n^{2-\gamma/2}\,, \\
 \Delta t_n &=& k_1\phi_n^{1-\gamma/2}\,,
\end{eqnarray}
where $k_1=\left(\case{2\sqrt{2\pi\gamma}\,\Gamma(1 +
1/\gamma)}{\Gamma(1/2 + 1/\gamma)}\right)\lambda^{-1/2}$ and
$k_2=c\left(\case{12\sqrt{2\pi}\,\Gamma(1/\gamma)}{\sqrt{\gamma}(2 +
\gamma)\Gamma(1/2 + 1/\gamma)}\right)\lambda^{-1/2}$. The solution to
these recurrence relations are the power-laws:
\begin{eqnarray}
 \phi_n &=& \phi_1 n^a, \quad a = -\frac{6c}{2(1+3c) +
 \gamma(1-3c)}\,,\\ \nonumber
 t_n &=& t_1 n^b, \quad b = \frac{(2+\gamma)}{2(1+3c) + \gamma(1-3c)}\,.
\end{eqnarray}
The energy density is
$\rho_\phi\sim\phi^\gamma\sim\,t^{a\gamma/b}$. In terms of the
cosmological scale-factor $R$ this can be written,
$\rho_\phi\sim\,R^{a\gamma/bc}$, i.e.
\begin{equation}
 \rho_\phi \sim R^{-6\gamma/(2+\gamma)}\,,
\end{equation}
independently of $c$ \cite{tur83}. The number of oscillations goes as
$n\sim\phi^{1/a}\sim\,t^{1/b}\sim\,R^{1/cb}$. Table~\ref{tabla} shows
the exponents of different quantities as functions of $n$, $R$ and $t$
for the cases of radiation- and matter- domination and for quadratic
and quartic potentials.

\begin{table}[b]
\begin{tabular}{|l|l|l|l|l|l|l|l|l|}
& $\phi (n)$ & $t (n)$ & $\rho_\phi (n)$ & $\phi (R)$ & $t (R)$ 
 & $\rho_\phi (R)$ & $n (t)$ & $n (R)$ \\ \hline
$c = 1/2~;~\gamma = 2$ & -3/4 & 1 & -3/2 & -3/2 & 2 & -3 & 1 & 2 \\ \hline
$c = 1/2~;~\gamma = 4$ & -1 & 2 & -4 & -1 & 2 & -4 & 1/2 & 1 \\ \hline
$c = 2/3~;~\gamma = 2$ & -1 & 1 & -2 & -3/2 & 3/2 & -3 & 1 & 3/2 \\ \hline
$c = 2/3~;~\gamma = 4$ & -2 & 3 & -8 & -1 & 3/2 & -4 & 1/3 & 1/2 \\
\end{tabular}
\bigskip
\caption{Power-law exponents for the evolution of field variables
during oscillations after inflation, for a quadratic ($\gamma=2$) and
quartic ($\gamma=4$) potential, assuming a radiation-dominated
($c=1/2$) and matter-dominated ($c=2/3$) universe.}
\label{tabla}
\end{table}

Let us return to the double well potential (\ref{poti}) which
interests us here. As we saw in the preceding Section, the quartic
term dominates at the end of inflation in the cases of interest, i.e
$\beta \gg1$. The subsequent evolution starting with $\phi=\phi_0$ at
$t=t_0$ goes as follows. The field oscillates in the quartic
term-dominated potential (where the mass term can be neglected),
decreasing in amplitude due to friction caused by the Hubble
expansion, until $V^{1/4}$ drops below the height of the barrier $h$
separating the two vacua, or in other words $\phi$ becomes of ${\cal
O}(v)$, and the field settles down in one of its potential minima. The
energy density in oscillations then decreases as for radiation
($\rho_\phi\sim\,R^{-4}$) independently of the rate of expansion. The
relation between the number of half-oscillations and the field
amplitude is $n\sim\phi^{-1}$ if the universe is radiation-dominated
and $n\sim\phi^{-1/2}$ if it is matter-dominated. Thus $\phi$ will be
found around one of its vevs $\pm\,v$ after having completed
$n=\tau_\r\chi_0$ half-oscillations in the radiation-dominated case
and $n=\tau_\m\sqrt{\chi_{0}}$ in the matter-dominated case, where
$\chi_0=\phi_0/v$, and $\tau_\r$, $\tau_\m$ are constants. (The
numerical values of these constants cannot be calculated given the
approximations we have made since they depend on the details of the
evolution towards the end of the oscillations (when the mass term
begins to be significant), and even more importantly on
$t_0\,(=c/H_\i)$, which sets the Hubble parameter at the beginning of
the oscillations and thus determines the friction term.) In general
$n$ is a function of $\chi_0$, $t_0$ and $\lambda$. However, rewriting
the equation (\ref{hubble}) using $\hat{\phi}=\phi/\phi_0$ and
$\hat{t}=\sqrt{\lambda}\phi_0\,t$, we have:
\begin{equation}
 \frac{\rmd^2\hat\phi}{\rmd\hat{t}^{2}} +
  3\frac{c}{\hat{t}}\frac{\rmd\hat\phi}{\rmd\hat{t}} + 
  \left(\hat\phi^3 - \frac{1}{\chi_0^2}\hat\phi\right) = 0\,.
\end{equation}
This equation has one parameter and the initial conditions are
$\hat\phi_0=1$, $\rmd\hat\phi/\rmd\hat{t}=0$. Therefore the number of
oscillations must depend on just $\chi_0$ and on
$\hat{t}_0=c\sqrt{\lambda}\phi_0/H_\i$. However, as we stated before,
if initially $H_\i\gg\omega=0.84\sqrt{\lambda}\phi_0$ the field
remains frozen. It only begins to oscillate when $H$ decreases below
$\omega$, i.e. at $\hat{t}_0\sim1$, hence the problem does not depend
on $\hat{t}_0$ or any other parameter as long as we are interested
only in the final state. We have checked numerically that $n$ does not
depend significantly on $\hat{t}_0$ when $\hat{t}_0\lesssim0.3$. In
the present case we have that the initial distribution of $\phi_0$ is
most probably concentrated around $\phi_0\sim0.1\,\lambda^{-1/4}H_\i$,
and $\hat{t}_0\lesssim0.1c\,\lambda^{-1/4}$, so there can be no
significant dependence on $\hat{t}_0$. (It may be that the initial
value of the Hubble parameter for the FLRW era, $H_0$, is somewhat less
than its inflationary value, $H_\i$, but even in this case the results
apply for small $\lambda$.) The proportionality constants can be
calculated numerically in this regime and are found to be
$\tau_\r=0.31$ and $\tau_\m=0.63$. When $\hat{t}_0$ exceeds unity the
Hubble parameter is smaller than $\omega$ at the beginning of the
oscillations and the reduction of the friction implies a proportional
increase of the number of oscillations:
$\tau_\r,\tau_\m\propto\hat{t}_0$ for $\hat{t}_0>1$.

The function $f(\chi_0)$ that equals $\pm1$ according to the value
$\phi=\pm\,v$ of the final state will then change sign just once as
$n$ increases by unity. Therefore we can write
\begin{eqnarray}
 f(\chi_0) &=& (-1)^{\rmint(\tau_\r \chi_0)}, \quad \for \quad 
  c = \frac{1}{2}\,, \nonumber \\
  &=&(-1)^{\rmint(\tau_\m \sqrt{\chi_0})}, \quad \for \quad 
  c = \frac{2}{3}\,, 
\label{fur}
\end{eqnarray}
where $\rmint(x)$ is the integer part of $x$. Small deviations from
these expressions occur for low values of $\chi_0$. The oscillations
start when
$t=t_*\simeq1/\sqrt{\lambda}\phi_0\simeq\lambda^{-1/4}H_\i^{-1}$, and
whether the universe is radiation- or matter- dominated at this time
is determined by whether $t_*$ is smaller or larger than the epoch of
matter and radiation equality $t_{\eq}$. (There is also the
possibility of a different expansion rate during the reheating
process.)

Note that the functions (\ref{fur}) apply only when the oscillations
begin and end during a period of expansion while $c$ is constant. As
an example of a more complex situation consider the case where the
oscillations start in the radiation-dominated era and end in the
matter-dominated one. According to Table~\ref{tabla} the amplitude of
the oscillations goes as $\phi\sim1/R$. The amplitude at the time of
matter-radiation equality is then
$\phi_\eq=\phi_0(t_*/t_\eq)^{1/2}\sim(\phi_0/t_\eq\sqrt{\lambda})^{1/2}$,
while the condition that the oscillations end in the matter-dominated
period is $\phi_\eq>v$. The number of oscillations in the
radiation-dominated period is just
$n_\r\simeq\phi_0/\phi_\eq=(t_\eq\sqrt{\lambda}\phi_0)^{1/2}$.
Therefore the number of half-oscillations completed in the
radiation-dominated period is proportional to $\sqrt\phi_0$. With
respect to the remaining oscillations that occur in the
matter-dominated period, since they start at $t_{\eq}$ well inside the
low friction regime, the constant $\tau_\m$ in Eq.(\ref{fur}) has to be
scaled by $\hat{t}_\eq=\sqrt{\lambda}\phi_{\eq}t_\eq$. Accordingly,
the matter-dominated period gives the number of oscillations
$n_\m\simeq\sqrt{\lambda}\phi_{\eq}t_\eq\sqrt{\chi_\eq}=
\lambda^{1/8}t_\eq^{1/4}\phi_0^{3/4}/\sqrt{v}$. Thus, for such mixed
situations, different functional dependencies of $\phi_0$ are expected
in the formula for the number of oscillations.

\section{Bias}
\label{bia}

The bias, defined as the difference in the probabilities of populating
the two discrete vacua, is given by the convolution
\begin{equation}
 b (\bar\chi) = \int f (\chi_0)\, P_\chi (\chi_0, \bar\chi)\, \rmd\chi_0\,.
\label{bias}
\end{equation}
The probability distribution of the field values at the end of
inflation $P_\chi(\chi_0,\bar\chi)$ is given by Eq.(\ref{prob})
(rewritten in terms of the normalized fields $\chi=\phi/v$ and
$\bar\chi=\bar\phi/v$), while the function $f(\chi_0)$ that gives the
sign of the field in the final vacuum state was obtained in the
previous Section for different cosmological situations. (Note that
$P_\chi(\chi_0,\bar\chi)$ is almost independent of spatial scale in
the FLRW era since observable scales correspond to only a few
e-foldings during inflation.) In the following we do a detailed
analysis of the bias function for the cases of radiation- and
matter-dominated universes and then present an approximation suitable
for more complicated situations.

The bias in the observable universe is well defined by Eq.(\ref{bias})
but since only the probabilistic distribution (\ref{dis}) is available
for $\bar\chi$, the predictions are also in terms of a probability
distribution for the bias which satisfies
\begin{equation}
 P_{b} = \sum P_{\bar\chi}\,\frac{\rmd\bar\chi}{\rmd b}\,,
\end{equation}
where $\bar\chi$ is understood as a function of $b$ (inverse of the
function (\ref{bias})), and the sum is over the different branches in
the solution of the equation $b=b(\bar\chi)$. The cumulative
probability for the bias to be less than some particular value is then
just the integral:
\begin{equation}
 P (|b|<x) = \sum\int_{b(\bar\chi)=-x}^{b(\bar\chi)=x}
 P_{\bar\chi}\,\rmd\bar\chi\,.
\end{equation}
The problem has basically two parameters,
\begin{equation}
 \alpha = \frac{\sigma}{v} \sim \frac{H_\i}{v}\,,
\label{defalpha}
\end{equation}
which gives the width of the Gaussian probability distribution
(\ref{prob}) in terms of the normalized field $\chi_0\equiv\phi_0/v$,
and $\beta\sim\lambda^{-1/4}\alpha$ (using Eq.(\ref{dis})) which gives
the width of the distribution (\ref{dis}) of the normalized average
field $\bar\chi\equiv\bar\phi/v$. The conditions we are assuming for
slow-roll $m\lesssim\,H_\i$ and $\lambda<1$ imply that
$\alpha<\min(\beta,\beta^2)$ but they do not constrain this parameter
to be greater or smaller than unity since we also require
$\beta\gg\,1$ (i.e. $H_\i\gg\,h$) in order for the field to be able to
jump the potential barrier during inflation.

\subsubsection{Radiation dominated universe}

In this case the number of half-oscillations $n\sim\tau_\r\,\chi_0$,
and $f(\chi_0)=(-1)^{\rmint(\tau_\r\,\chi_0)}$ is a square-wave. It
is easy to see that the bias is then a periodic function of
$\bar{\chi}$ with period $2\tau_\r^{-1}$, and can be expanded in the
Fourier series
\begin{equation}
 b (\bar\chi) = \frac{4}{\pi} \sum\limits_{n=0}^{\infty}
 \frac{\sin[(2n+1)\pi\tau_\r\bar\chi]}{(2n+1)} 
 \exp{\left[-(2n+1)\frac{\pi\tau_\r}{\sqrt{2}}\alpha\right]^2}\,, 
\label{aab}
\end{equation}
which converges exponentially fast. For $\alpha\gtrsim1$
(i.e. $H_\i\gtrsim\,v$) we can approximate the series by its first term
\begin{equation}
 b (\bar\chi) = \frac{4}{\pi} \sin(\pi\tau_\r\bar\chi)
 \exp{\left(-\frac{\pi\tau_\r}{\sqrt{2}}\alpha\right)^2}\,,
\label{bbc}
\end{equation}
showing that the bias is a sine function with an exponentially damped
amplitude. In the opposite case $\alpha\ll1$ (i.e. $H_\i\ll\,v$), more
terms of the series must be added so it converges to the Fourier
series for the square wave, i.e. identical to Eq.(\ref{aab}) without
the exponential factor. In this limit the Gaussian distribution
$P_{\chi}(\chi_0,\bar\chi)$ is appropriate inside the regions where
$f(\chi_0)$ has one sign or the other, but moving $\bar\chi$ from one
of these regions to another where $f(\chi_0)$ has {\em opposite} sign
causes the function $b(\bar\chi)$ to step as the error function,
$\erf(x)\equiv\case{2}{\sqrt{\pi}}\int_0^x{\rmd}t\,\rme^{-x^2}$.
Therefore for $\alpha\ll\,1$ we can write the periodic function $b$ in
the interval $[-\tau_\r^{-1},\tau_\r^{-1}]$ as
\begin{equation}
 b (\bar\chi) = \erf\left(\frac{1}{\sqrt{2}}\frac{\bar\chi}{\alpha}\right)
 - \erf\left[\frac{1}{\sqrt{2}}\frac{(\bar\chi+\tau_\r^{-1})}{\alpha}\right]
 - \erf\left[\frac{1}{\sqrt{2}}\frac{(\bar\chi-\tau_\r^{-1})}{\alpha}\right]\,.
\end{equation}
This function is very close to +1 or -1 except in the neighborhood of
the origin (or the points $n\tau_\r^{-1}$) where it is linearly
dependent on $\bar\chi$:
\begin{equation}
 b (\bar\chi) = \sqrt{\frac{2}{\pi}}\frac{\bar\chi}{\alpha}\,, \qquad
 \bar\chi < \alpha\,.
\end{equation}
In Fig.~\ref{juntos1}(a) we show the bias function for several values of
$\alpha$.

\begin{figure}[htb]
\epsfxsize\hsize\epsffile{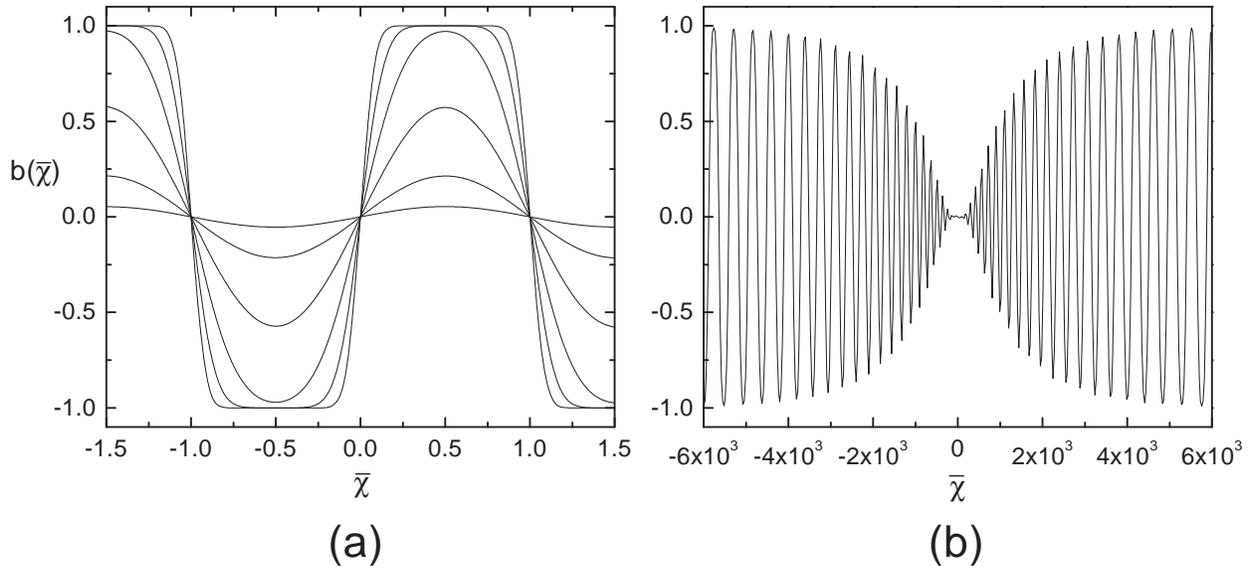}
\bigskip
\caption{The bias function for different values of $\alpha$ in (a) the
radiation-dominated and (b) the matter-dominated case. In the
left panel, the curves correspond, from top to bottom, to the values
$\alpha=0.05,0.1,0.2,0.4,0.6,0.8$, while on the right, the bias is
shown for $\alpha=100$.}
\label{juntos1}
\end{figure}

When $\beta\gg1$, we have that $\phi_0\sim\lambda^{-1/4}H_\i\gg\,v$,
and the initial values of $\bar\chi$ will be distributed with equal
probability in the interval $[-\tau_\r^{-1},\tau_\r^{-1}]$. Thus, in
this case there is no dependence on $\beta$, and the results are
relatively insensitive to the initial probability distribution for
$\bar\chi$, even if stochastic equilibrium is not achieved during
inflation. Therefore, the bias probability function $P(b)$ can be
simply calculated using
$P_{b}(b)=\frac{\tau_\r}{2}\frac{\rmd\bar\chi}{\rmd\,b}$ with
$\bar\chi$ in the interval $[0,\tau_\r^{-1}/2]$, yielding
\begin{eqnarray}
 P_{b} &=& \frac{1}{\pi} \left[\left(\frac{4}{\pi}\right)^2
\exp{(-\pi\tau_\r\alpha)^2} - b^2\right]^{-1/2}
 \quad \for \quad \alpha \gtrsim 1\,, \quad
|b|<\frac{4}{\pi}\exp{\left(-\frac{\pi\tau_\r}{\sqrt{2}}\alpha\right)^2}\,,
 \nonumber \\
 &=& \sqrt{\frac{\pi}{8}}\tau_\r\alpha
\quad \for \quad \alpha \ll 1\,,
\quad |b| \lesssim 1\,,
\end{eqnarray}
where we have shown the maximum value the bias can reach (see
Fig.\ref{juntos2}(a)). The cumulative probability for the bias is
$P(|b|<x)=2\tau_\r\bar\chi(x)$, where the function $\bar\chi(x)$ is
just the inverse of the bias function in the interval
$[0,\tau_\r^{-1}/2]$. This gives the approximate answer
\begin{eqnarray}
 P (|b|<x) &=& \frac{2}{\pi} \arcsin\left[\frac{\pi x}{4}
 \exp{\left(\frac{\pi\tau_\r}{\sqrt{2}}\alpha\right)^2}\right] 
 \quad \for \quad 
 x <\frac{4}{\pi}\exp{\left(-\frac{\pi\tau_\r}{\sqrt{2}}\alpha\right)^2}, 
 \quad \alpha \gtrsim 1\,,  \nonumber \\
 &=& \sqrt{2\pi}\tau_\r\alpha\,x \quad \for \quad x\lesssim 1, \quad \alpha \ll 1\,.
\label{prec}
\end{eqnarray}
This is plotted in Fig.~\ref{juntos2}(a) for various values of $\alpha$.

\begin{figure}[htb]
\epsfxsize\hsize\epsffile{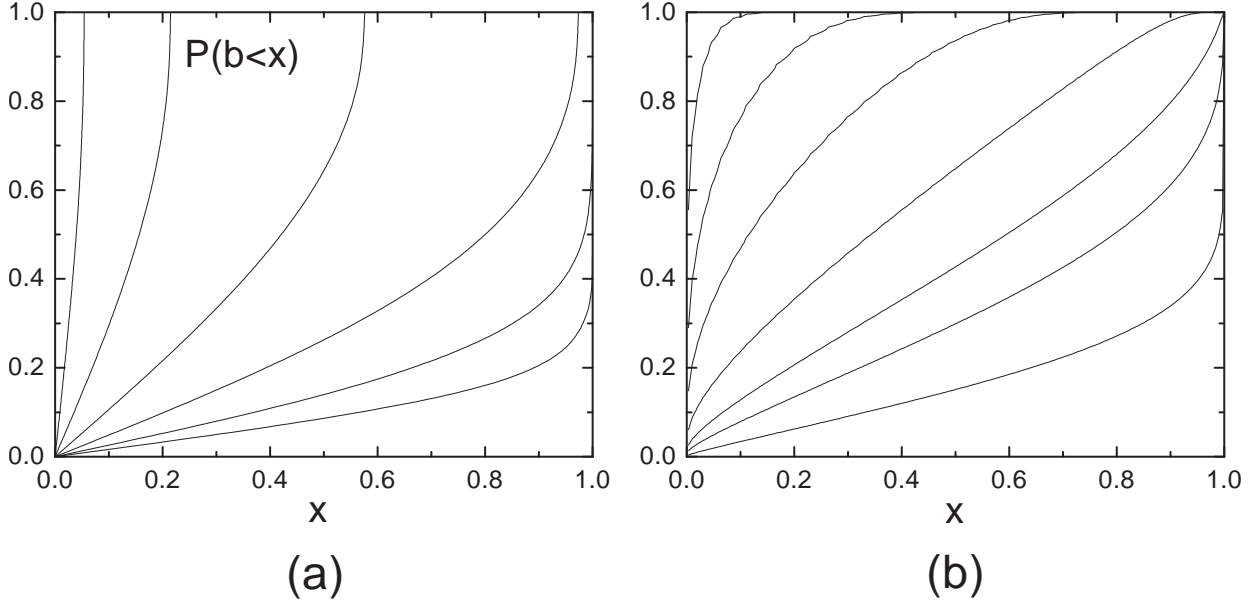}
\bigskip
\caption{The cumulative probability in (a) the radiation-dominated and (b) the matter-dominated case. In the left panel, the curves
correspond, from bottom to top, to the values $\alpha=0.05,0.1,0.2,0.4,0.6,0.8$ (as in Fig.~\protect\ref{juntos1}(a)),
while in the right panel the curves correspond, from bottom to top, to
the values $\alpha/\sqrt{\beta}=0.2,0.45,0.7,1.0,1.6,2.2,3.2$.}
\label{juntos2}
\end{figure}

\subsubsection{Matter dominated universe}

For the matter-dominated universe we have $n=\tau_\m\sqrt{\chi_0}$, and
the bias will not be a periodic function of $\bar\chi$. Consequently
the bias probability function is more sensitive to the initial
probability distribution of $\bar\chi$, in contrast to the
radiation-dominated case.  Taking the stochastic distribution as the
initial probability distribution for $\bar\chi$ the problem has two
parameters rather than just one as in the radiation-dominated case.

Of course, the bias function $b(\bar\chi)$ depends only on
$\alpha$. If $\alpha\ll1$ then the bias will be concentrated around
$\pm1$ since the width of the distribution will not allow different
final vacuum states to be reached. The function will be $\pm1$ except
near the transition points $\bar{\chi}_n=\tau_\m^{-2}n^2$, where it
jumps as $\pm\erf(\case{\bar\chi-\bar\chi_n}{\sqrt{2}\alpha})$. In
fact this will also be the case for values of $\bar\chi$ greater than
$\alpha^2$ (with $\alpha\gtrsim1$), because at this point the spacing
of $\chi_0$ which results in different final states is
$\gtrsim\alpha$. In the opposite case, $\bar\chi\lesssim\alpha^2$,
significant cancellation take place and the bias will be reduced. An
excellent approximation in this regime (if $\alpha\gtrsim1$) is given
by the formula (\ref{bbc}), if we allow for variations in the period
and the amplitude of the sine function according to the effective
variation of $\alpha$ with $\bar\chi$ (taking into account the
increasing distance between the $\bar\chi_n$ with $n$ in the matter
dominated case), i.e.
\begin{equation}
 b (\bar\chi) = \frac{4}{\pi} \sgn(\bar\chi) \exp\left[-\frac{1}{|\bar\chi|}
 \left(\frac{\pi\tau_\m\alpha}{2\sqrt{2}}\right)^2\right] 
\sin\left(\pi\tau_\m\sqrt{|\bar\chi|}\right)\,, \quad \for \quad
\bar{\chi} \lesssim \alpha^2\,. 
\label{bf}
\end{equation}
In Fig.~\ref{juntos1}(b) we show an example of the bias function for the
matter-dominated case.

For the case of interest, i.e. $\beta\gg1$, the mass term can be
neglected in Eq.(\ref{dis}) and the initial probability for $\bar\chi$
writes
\begin{equation}
 P_{\bar\chi} = \frac{\sqrt{\pi}}{2^{3/4}3^{1/4}\Gamma(5/4)\beta}
 \exp\left(-\frac{2\pi^2}{3}\beta^{-4}\bar\chi^4\right)\,.
\end{equation}
This function is practically constant for $\bar\chi<\beta/2$ and then
falls until $\bar\chi=\beta$ where it is nearly zero. Thus for
$\beta\gg\alpha^2$ the bias is $\pm1$ with high probability because
$\bar\chi\sim\beta$. In this case we recover the linear
behavior (\ref{prec}) of the cumulative probability for the bias in
the radiation-dominated case, but now the spacing $\tau_\r^{-1}$ depends
on $\beta$:
\begin{equation}
 P (|b|<x) = \sqrt{2\pi}\frac{\alpha}{0.71\sqrt{\beta}}\,x\,,
  \quad \for \quad x\lesssim 1,\quad \alpha \ll \sqrt{\beta}\,,
\end{equation}
where the precise coefficient in the identification
$\tau_\r^{-1}\sim\sqrt{\beta}$ is
$\case{2^{5/8}\Gamma(9/8)}{\sqrt{5}\Gamma(5/4)}\simeq0.71$, as can
be calculated from the average value of the inverse spacing.

For $\alpha<\beta\lesssim\alpha^2$ the bias has higher probability
between $-1$ and $1$. In this case the probability $P(|b|<x)$ is
dominated by the sector in the bias function (\ref{bf}) where
$x>\case{4}{\pi}\exp\left[-\case{1}{|\bar\chi|}
\left(\case{\pi\tau_\m\alpha}{2\sqrt{2}}\right)^2\right]$ and thus
$|\bar\chi|<\bar\chi_x= 
-(\case{\pi\tau_\m\alpha}{2\sqrt{2}})^2/\log(\case{\pi}{4}x)$. Then
\begin{equation}
 P (|b|<x) = 2\int_0^{\bar\chi_x} P_{\bar\chi}\,\rmd\bar\chi = 1 - 
  \frac{\Gamma(\case{1}{4},\beta^{-4}\bar\chi_{x}^4)}{\Gamma(\case{1}{4})}\,,
  \quad \for \quad \sqrt{\beta} \lesssim \alpha\,,  
\label{ff}
\end{equation}
where the function $\Gamma(
x,y)=\int_y^{\infty}t^{x-1}\rme^{-t}\rmd\,t$ is the incomplete gamma
function. The behavior of the cumulative probability function changes
from linear to the form (\ref{ff}) as $\beta/\alpha^2$ becomes smaller
and in both limits it depends only on this combination of the
parameters as shown in Fig.~\ref{juntos2}(b).

Let us recapitulate the main features. In the radiation-dominated case
there is no dependence on $\beta$, and the probability of a small bias
$b$ is just $\sim\,b\alpha$ for $\alpha\sim\,H_\i/v\ll1$, while for
$\alpha\gtrsim1$ it increases exponentially towards unity as
$\sim\,b\rme^{\alpha^2}$. For the matter-dominated universe, the
parameter that regulates the bias probability and plays the same role
as $\alpha$ in the radiation-dominated case is
$\alpha/\sqrt{\beta}\sim\sqrt{\lambda^{1/4}H_\i/v}$. If
$\beta\equiv\lambda^{-1/4}H_\i/v\gg\alpha^2$ the probability of a
small bias is also small, $\sim\,b\alpha/\sqrt{\beta}$, but if
$\beta\lesssim\alpha^2$ it increases exponentially fast, being near
unity when $b\rme^{\alpha^2/\beta}\sim 1$.

In a general situation where the number of half-oscillations $n$ is
some function $n(\phi_0)$, the change in the initial field necessary
to change the number of half-oscillations by unity is
\begin{equation}
 \Delta \phi_0 = \left[\frac{\rmd n(\phi_0)}{\rmd\phi_0}\right]^{-1}\,.
\end{equation}
This needs to be compared with the width of the initial distribution
$\sigma\simeq\,H_\i$ in order to estimate the bias. If $H_{\i}{\rmd
n(\phi_0)/\rmd\phi_0}\gg1$ the bias will be exponentially damped and
the domain wall network will be stable. On the contrary, if
$H_{i}{\rmd n(\phi_0)/\rmd\phi_0}\ll1$ the bias will be of order unity
and the subsequent evolution will make the wall network collapse
exponentially fast \cite{clo96,hin96,lsw97}.

\section{History of the field energy density and wall formation}
\label{hist}

The history of the energy density of the coherent field component for
the $Z_2$-symmetry breaking potential (\ref{poti}) is approximately as
follows. During inflation the energy density in the $\phi$ field is of
${\cal O}(H_\i^4)$, and the value of the field is
$\phi_0\sim0.3\lambda^{-1/4}H_\i$ in stochastic equilibrium. (We have
assumed that the quartic term in the potential dominates so that
$H_\i>h$, where $h$ is the height of the potential barrier; this is a
necessary condition for inflation to produce domain walls, since
otherwise the field will be settled in one of the two minima both
during and after inflation.) If the symmetry is restored by thermal
effects following reheating after inflation then domain walls will
form again by the Kibble mechanism \cite{kib76}. However if the field
is sufficiently weakly coupled this will not happen \cite{lowscale}
and $H_\i<h$ will then be a sufficient condition for inflation to
solve the wall problem.\footnote{In principle symmetry restoration can
also occur through non-thermal effects during ``preheating'' leading
to the formation of topological defects \cite{pre} --- however
according to the results of Ref.\cite{pre2} this will not happen for
the model under consideration here.}

After inflation, the field perturbations do not evolve significantly
until $H$ becomes less than $\omega\sim\lambda^{1/2}\phi_0\sim
0.3\lambda^{1/4}H_\i$. Subsequently the field starts oscillating. At
this point its energy density is still $\sim0.01H_\i^4$ while the
energy density of the universe is $\sim3M_\P^2\,H^2\sim
0.3\lambda^{1/2}H_\i^2\,M_\P^2$. Thus for the field $\phi$ not to
dominate the energy density we require
$\lambda^{1/2}M_\P^2>0.03H_\i^2$, i.e. $\phi_0$ should not
significantly exceed $M_\P$, which seems natural. If
$\lambda<10^{-3}(H_\i/M_\P)^4$ the stationary stochastic distribution
for the initial conditions will not apply.

The energy density of the oscillating field redshifts as radiation
hence its relative energy, compared with the total, either decreases
or remains constant depending on whether the universe is
matter-dominated or radiation-dominated. Therefore the energy density
of the field can {\em never} dominate since it did not do so initially
when the field was began to oscillate at the end of inflation. However
after the domain walls form, the relative energy density starts
increasing again (at least until the wall network decays due to the
bias). The field is released when $t=t_*\sim(\lambda^{1/4}H_\i)^{-1}$
and subsequently its energy density decays as
$\rho_\phi=(R_*/R)^4H_\i^4$. The walls will form when $\rho_\phi$
decreases below $h^4$ i.e. at a time $t_\w$ determined by
\begin{equation}
 \frac{R(t_\w)}{R(t_*)} \sim \frac{H_\i}{h}\,.
\end{equation}

Using the results of the preceding sections we show in
Fig.~\ref{final} the outcome for wall formation and survival in the
parameter space of the scalar field model (\ref{poti}), for two
specific set of inflationary parameters --- (a)
$H_\i=2\times10^{4}$~GeV corresponding to an inflationary scale of
$\sim10^{11}$~GeV, $T_\reh=10^3$~GeV, and (b) $H_\i=10^{15}$~GeV
corresponding to an inflationary scale of $\sim10^{16}$~GeV,
$T_\reh=10^9$~GeV. Reheating is assumed to occur while the inflaton
oscillates in a quadratic potential, so the scale-factor evolves as
for a matter-dominated universe and reheating ends at
$t_\reh\simeq\,M_\P/T_{\reh}^2$.

\begin{figure}[htb]
\epsfxsize\hsize\epsffile{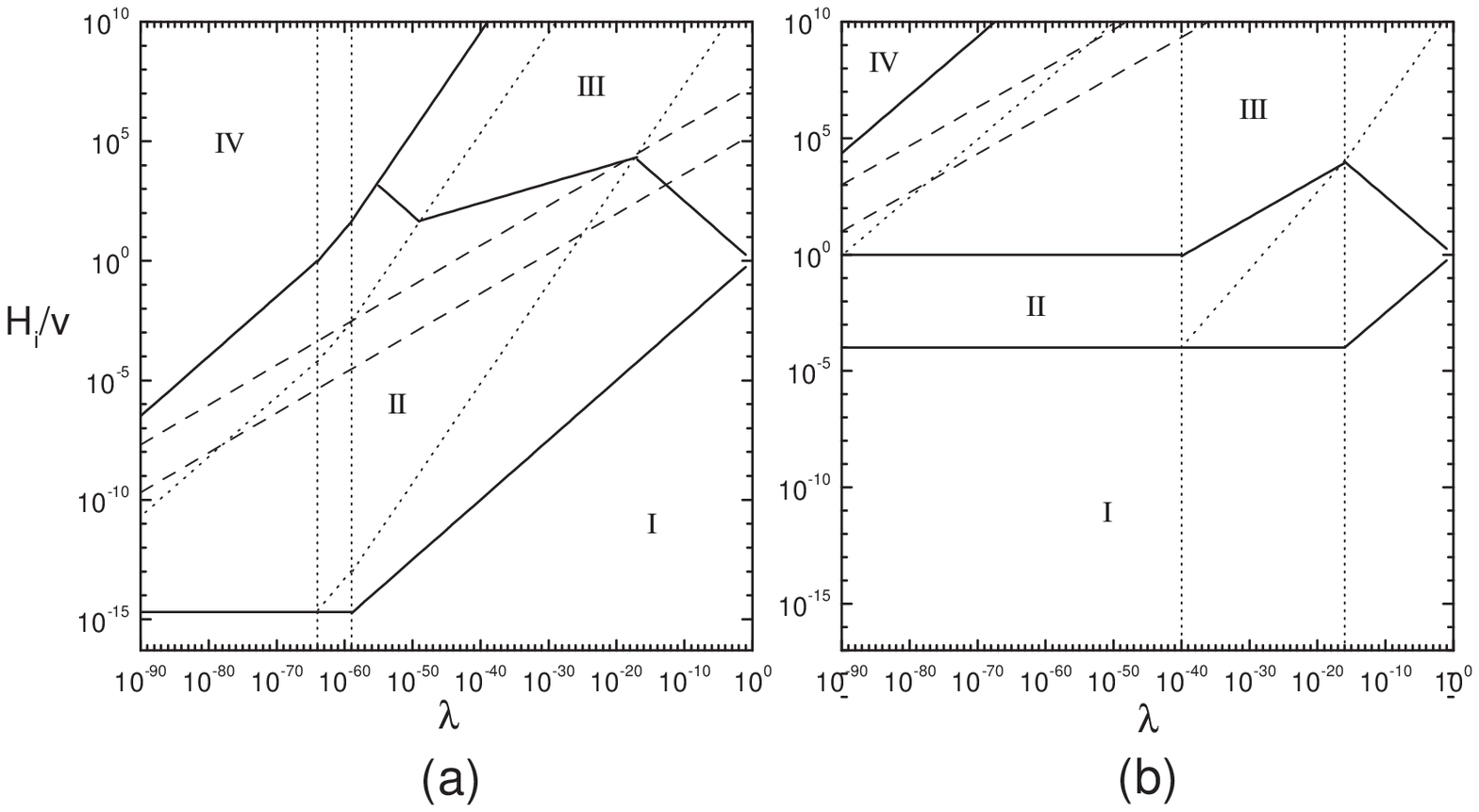}
\bigskip
\caption{The schematic situation concerning wall formation in the
parameter space of the model (\protect\ref{poti}) discussed in this
work. In region I the inflationary perturbations are too small to jump
the potential barrier and the symmetry remains always broken. In
regions II and III walls form with, respectively, high and low
bias. In region IV the walls have not yet formed by the present time
and the field is still oscillating. The walls form in the
matter-dominated era above the upper dotted line, during the reheating
epoch below the lower dotted line, and during the radiation-dominated
era in the region in between. The left vertical dotted line separates
regions where the oscillations begin at the reheating epoch (right
side) or in the radiation-dominated era (left side). The right
vertical dotted line separates regions where stochastic equilibrium
applies (right side) or where different initial conditions apply (we
have conservatively assumed $\phi_0\sim\,M_\P$ for the initial
condition in this region). The dashed lines indicate the domain wall
tension $\sigma$ corresponding to wall domination at present
($\sigma^{1/3}\simeq100$~MeV, lower line), and excessive CMB
anisotropy ($\sigma^{1/3}\simeq1$~MeV, upper line). Above these lines
the tension is low enough to avoid conflict with observations. In the
neighborhood of the line separating regions II and III, the bias takes
values between zero and unity. In region III it approaches zero
exponentially fast so the wall system survives. In region II the bias
is large and the wall network collapses so there is no cosmological
domain wall problem. The inflationary parameters are (a)
$H_\i=2\times10^{4}$~GeV, $T_\reh=10^3$~GeV, and (b)
$H_\i=10^{15}$~GeV, $T_\reh=10^9$~GeV.}
\label{final}
\end{figure}

\section{Discussion}

We have developed the tools for computing the bias in domain wall
formation in a specific Higgs-like model with a $Z_2$-symmetry. As a
general rule we find that the quantity which controls the bias is
$H_\i\,(\rmd\,n(\phi_0)/\rmd\phi_0)$, where the function $n(\phi_0)$
gives the total number of half-oscillations performed by the field,
starting with the value $\phi_0$ at the end of inflation, until it
settles in one of its symmetry-breaking minima. This has to be
evaluated taking into account both the form of the scalar potential
and the expansion history of the universe. An exponentially small bias
corresponds to $H_\i\,(\rmd\,n(\phi_0)/\rmd\phi_0)\gtrsim1$ while the
bias increases approximately linearly for
$H_\i\,(\rmd\,n(\phi_0)/\rmd\phi_0)<1$. To obtain the probability
distribution of the bias, one needs the initial probability
distribution for $\phi_0$. We have used the stochastic description
\cite{sta82} of the field fluctuations during inflation to compute
this.

The results of our detailed study for the Higgs-like potential are
shown in Fig.~\ref{final} adopting both a high ($10^{16}$~GeV) and an
intermediate ($10^{11}$~GeV) energy scale for inflation. We see that
stable domain wall formation does not occur in most of the parameter
space. For intermediate-scale inflation, the region where a problematic
stable domain wall network forms is much smaller than for GUT-scale
inflation and the domain wall problem is practically eliminated in
this case.

The results of this paper can be applied to other models as well. For
a harmonic potential as in the case of an axion field we recover the
results of Ref.\cite{ll90}. For a periodic potential the distribution
of initial values of $\bar{\phi}$ is not relevant and we have a
situation similar to the case of a Higgs-like potential in a
radiation-dominated universe which was studied in
Section~\ref{bia}. The parameter that controls the bias in this case
is $H_\i/f$ where $f$ is the period of the potential (i.e. the scale
of Peccei-Quinn symmetry breaking for the axion field).

In Ref.\cite{dgs01} the authors consider domain wall formation in a
model with a Higgs-like potential for a dilatonic-type scalar field
with a {\em very} small coupling constant, $\lambda\sim10^{-88}$, and
a vacuum expectation value $v<10^{11}$~GeV. As the authors
acknowledge, the issue of domain wall formation is somewhat subtle in
this extremely weakly coupled theory. For the model considered in the
present work, such couplings imply wall formation around the present
epoch or later (see Fig.\ref{final}), so they would not in fact be
astrophysically relevant. Furthermore the walls are formed in the high
bias region unless $H_\i\gtrsim10^{13}$~GeV. However unlike the model
considered in this work, the dilatonic field of Ref.\cite{dgs01} also
couples universally with matter through the term
\begin{equation}
 {\cal L}_{\rm int} = \exp\left(\frac{\phi}{M^*}\right){\theta^\mu}_\mu ,
\label{aaa}
\end{equation}
where $\theta^\mu_\mu$ is the trace of the energy-momentum tensor and
${M^*}^2\gtrsim(10^3-10^4)M_\P^2$.  During inflation the trace
$\theta^\mu_\mu$ is non-zero, driving the field quickly to large
negative values $\phi\lesssim-70M^*$ (where the exponential factor in
Eq.(\ref{aaa}) makes the size of this interaction term comparable to
the Higgs-like potential). Thus the effective mass is much smaller than
$H_\i$ and the generation of fluctuations of size $H_\i$ is
unavoidable. After inflation the term (\ref{aaa}) decreases rapidly,
so the field must relax in the quartic potential alone, starting at
scales higher than the Planck mass. As we have previously discussed
the absence of a force term strong enough to drive the field to the
origin will make the potential energy of the field dominate the energy
density of the universe.  In Ref.~\cite{dgs01} the authors also
introduce a term that couples the field coherently to the thermal bath
during the radiation-dominated era,
\begin{equation} 
 V_\therm = \frac{\kappa}{2} H^2 \phi^2 \sim \frac{T^4}{M_\P^2} \phi^2\,,
\label{bbb} 
\end{equation}
where $\kappa$ is a numerical constant. The intention in doing so is
to drive the field to the origin, restoring the symmetry, so domain
walls would be formed when $H$ falls below the (vacuum) mass of the
field. However, if $\phi$ is initially greater than the Planck mass
this term will exceed the energy density of the universe so the
treatment is not consistent. This problem does not occur if
$\kappa\ll1$ but in that case the mass of the field is {\em always}
much smaller than $H$ so the thermal term cannot affect the field.

Apart from this problem with the initial conditions, it is interesting
to examine the effect of the coupling (\ref{bbb}) with the thermal
bath on the  damping of the oscillations. The equation of motion for the field
in the radiation-dominated universe with such a potential term is
exactly solvable, with the general solution
\begin{equation}
 \phi(t) = c_1 t^{-(1+\sqrt{1-4\kappa})/4} + 
           c_2 t^{-(1-\sqrt{1-4\kappa})/4} .
\label{ttd}
\end{equation}
For $\kappa\ll1/4$ the friction dominates and the field decreases very
slowly as $t^{-\kappa/2}$ (dominant term in Eq.(\ref{ttd})), while for
$\kappa>1/4$ the behavior is oscillatory and the amplitude decreases
as $t^{-1/4}\sim\,R^{-1/2}$. Thus in either case the field amplitude
decreases more slowly than for the quartic potential we have
considered, where $\phi\sim\,R^{-1}$, or even for the quadratic
potential, where $\phi\sim\,R^{-3/2}$. This makes it even more
unlikely that the walls can be formed before the present epoch. Thus
this interesting attempt to do away with dark matter in galaxies by
modifying the gravitational force law cannot work.

\acknowledgements{We would like to thank Graham Ross for helpful
comments and encouragement. HC was supported by a CONICET Fellowship
(Argentina).}

\end{document}